\documentclass[reprint,amsmath,amssymb,aps,superscriptaddress,nofootinbib,twocolumn,10pt]{revtex4-1}
\usepackage[colorlinks,linkcolor=blue,citecolor=blue,urlcolor=blue]{hyperref}
\usepackage{graphicx,epstopdf}
\usepackage{dcolumn}
\usepackage{subfigure}
\usepackage{bm}
\usepackage{amsmath,amsthm,amsfonts,amssymb,graphicx,multirow,dcolumn,bm,latexsym,soul,nicefrac}
\usepackage{acronym}
\usepackage{enumitem}
\usepackage{array}
\usepackage{multirow}
\usepackage{appendix}
\usepackage{longtable}
\usepackage{threeparttable}
\newacro{GW}{gravitational-wave}
\newacro{GR}{general relativity}
\newacro{PTA}{Pulsar Timing Array}
\newacro{SGWB}{stochastic gravitational-wave background}
\newacro{LIGO}{Laser interferometry Gravitational-Wave Observatory}
\newacro{TDI}{Time Delay Interferometry}
\newacro{TQ}{TianQin}
\newacro{CO}{Compact Object}
\newacro{LISA}{Laser Interferometry Space Antenna}
\newacro{CBC}{compact binary coalescence}
\newacro{BH}{black hole}
\newacro{SBBH}{stellar-mass binary black hole}
\newacro{PN}{post-Newtonian}
\newacro{PA}{post-adiabatic}
\newacro{PSD}{power spectral density}
\newacro{ORF}{overlap reduction function}
\newacro{PLIS}{power-law integrated sensitivity}
\newacro{FIM}{Fisher information matrix}
\newacro{SNR}{signal-to-noise ratio}
\newacro{EMRI}{Extreme Mass Ratio Inspiral}
\newacro{MBH}{Massive Black Hole}
\newacro{LSO}{Last Stable Orbit}
\newacro{SF}{Self-Force}
\newacro{BHPT}{Black Hole Perturbation Toolkit}
\newacro{BHPC}{Black Hole Perturbation Club}
\newacro{FEW}{FastEMRIWaveforms}
\newcommand{\TRC}{MOE Key Laboratory of TianQin Mission,
TianQin Research Center for Gravitational Physics \& School of Physics and Astronomy,
Frontiers Science Center for TianQin,
Gravitational Wave Research Center of CNSA,
Sun Yat-sen University (Zhuhai Campus), Zhuhai 519082, China.}

\begin{document}

\title{A Deep Learning Framework for Amplitude Generation of Generic EMRIs}

\author{Yan-bo Zeng}
\affiliation{\TRC}
\author{Jian-dong Zhang}
\email{zhangjd9@mail.sysu.edu.cn}
\affiliation{\TRC}
\author{Yi-Ming Hu}
\affiliation{\TRC}
\author{Jianwei Mei}
\affiliation{\TRC}

\begin{abstract}
One of the main targets for space-borne gravitational wave detectors is the detection of \acp{EMRI}.
Data analysis of \acp{EMRI} requires waveform models that are both accurate and fast.
The major challenge for the fast generation of such waveforms is the generation of
the Teukolsky amplitudes for generic (eccentric and inclined) Kerr orbits.
The requirement for modeling ${\sim 10^5}$ harmonic modes in
a four-dimensional parameter space makes traditional approaches,
including direct computation or dense interpolation, computationally prohibitive.
To overcome this issue, we introduce a convolutional encoder-decoder architecture
for a fast and end-to-end global fitting of the Teukolsky amplitudes.
We also adopt a transfer learning strategy to reduce the size of the training dataset,
and the model is gradually trained
from the simplest Schwarzschild circular orbits to generic Kerr orbits step by step.
Within this framework,
we obtain a surrogate model based on a semi-analytical Post-Newtonian dataset,
and the full harmonic amplitudes can be generated within milliseconds,
while the median mode-distribution error for generic orbits is $\sim 10^{-3}$.
This result indicates that the framework is viable for the construction of efficient waveform models for \acp{EMRI}.
\end{abstract}

\maketitle

\section{Introduction}
When a stellar-mass \ac{CO} moves around a \ac{MBH} within the strong-field regime,
it will emit \ac{GW} and inspiral into the \ac{MBH}.
This kind of \ac{EMRI} system is one of the major targets for future space-borne gravitational-wave observatories such as TianQin~\cite{TianQin:2015yph,Fan:2020zhy,TianQin:2020hid} and \ac{LISA}~\cite{Amaro-Seoane:2007osp, Amaro-Seoane:2012lgq, Babak:2017tow, Berry:2019wgg, LISA:2024hlh}.
The small mass ratio dictates a slow inspiral that can include ${\sim 10^5}$ cycles within the detector's sensitive band.
Thus, the emitted \acp{GW} offer a sensitive probe for astrophysics~\cite{Bonga:2019ycj,Fan:2024nnp}, cosmology~\cite{MacLeod:2007jd,Zhu:2024qpp}, and high-precision tests of \ac{GR}~\cite{Gair:2012nm, Maselli:2020zgv, Zi:2021pdp, Zi:2022hcc}.
However, in order to achieve these objectives, an efficient data analysis method~\cite{Zhang:2022xuq,Ye:2023lok} is needed to search the signals and estimate their parameters. 
The first step is to obtain a precise and accurate waveform model that can be generated rapidly~\cite{Gair:2004iv, MockLISADataChallengeTaskForce:2009wir, Barack:2003fp, Cutler:2007mi, Hinderer:2008dm, Miller:2020bft,LISAConsortiumWaveformWorkingGroup:2023arg}.

Gravitational self-force (GSF) theory provides the most rigorous framework for modeling EMRIs,
by treating the system as a perturbative expansion
on the background of the central Kerr \ac{BH} according to the mass ratio $\epsilon$
\cite{Mino:1996nk,Quinn:1996am,Gralla:2008fg,Pound:2009sm,Detweiler:2011tt,Pound:2012nt,Gralla:2012db,Harte:2014wya,Poisson:2011nh,Barack:2018yvs,Pound:2021qin}.
The validity of the GSF framework has been clearly demonstrated through comparisons with full numerical relativity simulations, showing strong agreement even at moderate mass ratios~\cite{Wardell:2021fyy} and during the plunge stage~\cite{Compere:2021zfj,Kuchler:2024esj,Honet:2025dho}.
At the leading adiabatic order, the inspiral is governed by the dissipative part of the first-order self-force, which is equivalent to the energy and angular momentum fluxes of the emitted \ac{GW}.
Consequently, within the osculating geodesic framework, the orbit can be modeled as a sequence of geodesic snapshots~\cite{Hinderer:2008dm, Hughes:2005qb} whose slow evolution is directly driven by these fluxes. However, higher-order treatments require a more systematic self-force description~\cite{Pound:2021qin}, and methods such as the near-identity transformation and multiscale expansions are needed to separate the dissipative evolution and the osculating part.

Because these fluxes are constructed by summing individual harmonic modes obtained from the Teukolsky equation, computing these amplitudes forms the central computational bottleneck in \ac{EMRI} waveform generation~\cite{Hughes:2021exa, LISAConsortiumWaveformWorkingGroup:2023arg}.
While various numerical frequency-domain solvers~\cite{Nakamura:1987zz, Shibata:1993uk, Kennefick:1998ab, Glampedakis:2002ya, Drasco:2005kz, Fujita:2004rb, Fujita:2009us,Yin:2025kls} and semi-analytical \ac{PN} methods~\cite{Mino:1997bx, Sasaki:2003xr, Sago:2005fn, Ganz:2007rf, Fujita:2012cm, Shah:2014tka, Fujita:2020zxe}---along with modern toolkits~\cite{BHPToolkit, Nasipak:2022xjh, Nasipak:2023kuf, Nasipak:2025tby}---have been developed to solve this problem, the challenge remains significant.
Generally speaking, this amplitude bottleneck manifests in two distinct but related aspects.

The first is the high cost of offline data generation. 
The construction of a complete waveform model requires dense sampling of the four-dimensional parameter space $(a, p, e, x_I)$.
However, generating the data for even a single point is already a hard task.
The rich tri-periodic nature of a generic Kerr orbit excites a spectrum of ${\sim 10^5}$ harmonic modes~\cite{Drasco:2005kz, Hughes:2021exa},
and calculating the amplitude of each mode involves solving the Teukolsky equation.
The total computational effort also scales with orbital complexity;
for reference, a full 1st order self-force calculation is estimated to cost ${\sim 10^4}$ CPU-hours for a single generic orbit~\cite{vandeMeent:2017bcc}.

The second problem emerges in the online generation.
The initial version of \ac{FEW} for Schwarzschild orbits~\cite{Chua:2020stf, Katz:2021yft} employed Roman to predict coefficients on a compressed basis~\cite{Chua:2018woh}, which, in theory, offered a path of high efficiency and accuracy for high-dimensional problems. 
However, in practice, this approach struggled to accurately fit weaker modes and lacked flexibility for adaptive mode selection, which is a critical requirement for the analysis of high-SNR EMRI signals.
Consequently, the recent extension to Kerr equatorial orbits~\cite{Chapman-Bird:2025xtd} transitioned to a traditional interpolation.
This approach successfully solved the precision and flexibility issues but also introduced a memory wall.
It has been estimated that extending it to the ${\sim 10^5}$ modes of a generic inspiral would require over 500 GB of memory for the spline coefficients, rendering it impractical for current hardware.
\ac{EMRI} waveform modeling thus faces a tension between efficiency and accuracy.
A fast, accurate, and scalable method for obtaining the full harmonic content for generic Kerr orbits thus remains an open problem.

To address this, we propose a global fitting surrogate model built upon a convolutional encoder-decoder architecture for mode prediction. 
Our framework treats the full set of harmonic amplitudes as a structured tensor,
and the indices of this tensor correspond to the harmonic mode indices. 
This perspective allows us to employ a convolutional decoder to explicitly learn these local relationships, analogous to how convolutional networks process spatial information in images.
By directly learning an end-to-end mapping from the 4D orbital parameters to this amplitude tensor, our convolutional decoder can explicitly model correlations between adjacent modes.
Our methodology also includes a transfer learning strategy structured as a curriculum.
The model is first trained on the simplest orbital configurations and is then progressively exposed to more complex physics.
Thus the required training-dataset size could be reduced. 
This paper uses this framework to test the viability and scalability of a global fitting paradigm for the EMRI amplitude problem, with the PN dataset serving as a challenging, high-dimensional testbed.

This paper is organized as follows.
In Sec.~\ref{sec:waveform}, we review the theoretical background of EMRI waveform modeling, including orbital geometry and the Teukolsky formalism.
In Sec.~\ref{sec:methodology}, we introduce our proposed deep learning framework, covering the encoder-decoder architecture and the curriculum-based transfer learning strategy.
The numerical results and performance benchmarks across various orbital classes are presented in Sec.~\ref{sec:result}.
Finally, we summarize our findings and discuss future research directions in Sec.~\ref{sec:conclusion}.

\section{Waveform modeling of EMRI} \label{sec:waveform}

\subsection{Orbital Geometry and Classification} \label{sec:orbit_geometry}

In an \ac{EMRI} system, the dynamics of the secondary are governed by geodesic motion in the Kerr spacetime up to leading order. The integrability of these geodesics is ensured by three constants of motion: the energy $E$, axial angular momentum $L_z$, and Carter constant $Q$~\cite{Carter:1968rr}. This complete set of integrals allows any bound geodesic to be uniquely specified by a set of three orbital parameters. Following common convention, we use a quasi-Keplerian set: the semi-latus rectum $p$, eccentricity $e$, and the cosine of the inclination angle $x_I = \cos\iota$~\cite{Hughes:1999bq, Schmidt:2002qk}. Together with the primary black hole's dimensionless spin $a$, these four parameters, $(a, p, e, x_I)$, define the four-dimensional parameter space that fully specifies an orbital configuration, or ``snapshot''.

The parameters $a$, $e$, and $x_I$ define three fundamental binary properties that allow for a systematic classification of orbital geometries:
\begin{itemize}[noitemsep,topsep=0pt]
    \item \textbf{Spacetime:} The background is either \textit{Schwarzschild} ($a=0$) or \textit{Kerr} ($a \neq 0$).
    \item \textbf{Eccentricity:} The orbit is either \textit{circular} ($e=0$) or \textit{eccentric} ($e>0$).
    \item \textbf{Inclination:} The orbital plane is either \textit{equatorial} ($|x_I|=1$) or \textit{inclined} ($|x_I|<1$). We adopt the convention where $x_I \in [-1, 1]$ uniformly describes both prograde ($x_I > 0$) and retrograde ($x_I < 0$) orbits~\cite{Hughes:1999bq, Chapman-Bird:2025xtd}.
\end{itemize}

Specific combinations of these properties yield a hierarchy of distinct orbital classes, characterized by decreasing symmetry and increasing complexity. Our work, particularly our training strategy, is structured around this hierarchy. We define the specific categories in Table~\ref{tab:orbit_classification}.

\begin{table}[!htbp]
\caption{Classification of orbital geometries based on the parameters $(a, e, x_I)$. These categories form the basis of our curriculum learning strategy. It should be noted that the ``inclined circular'' orbit is also called ``spherical'' orbit in some literature, e.g., Refs.~\cite{Wilkins:1972rs, Teo:2020sey}.}
\label{tab:orbit_classification}
\begin{ruledtabular}
\begin{tabular}{llc}
\textbf{Orbital Type} & \textbf{Notation} & \textbf{Constraints} \\
\hline
Schwarzschild Circular & SC & $a=0, e=0, x_I=1$ \\
Schwarzschild Eccentric & SE & $a=0, e>0, x_I=1$ \\
Kerr Equatorial Circular & KEC & $a\neq0, e=0, |x_I|=1$ \\
Kerr Equatorial Eccentric & KEE & $a\neq0, e>0, |x_I|=1$ \\
Kerr Inclined Circular & KIC & $a\neq0, e=0, |x_I|<1$ \\
Kerr Generic & KG & $a\neq0, e>0, |x_I|<1$ \\
\end{tabular}
\end{ruledtabular}
\end{table}

\subsection{Adiabatic inspiral waveform construction}
The integrability of Kerr geodesics ensures that bound orbits are tri-periodic.
This motion is also characterized by three fundamental frequencies—$\Omega_r$, $\Omega_\theta$, and $\Omega_\phi$—corresponding to the radial, polar, and azimuthal components of the orbit, respectively, as measured in Boyer-Lindquist coordinate time $t$~\cite{Schmidt:2002qk}.
Consequently, the emitted gravitational-wave strain $h=h_+-ih_\times$ decomposes into a discrete sum over harmonic modes.
Each mode is indexed by integers $(\ell, m, n, k)$, where $(\ell, m)$ label the spin-weighted spheroidal harmonics, and the harmonic numbers $(n, k)$ correspond to the radial and polar motions~\cite{Drasco:2005kz, Hughes:2021exa}.
The frequency of each mode, $\omega_{mnk}$, is a linear combination of these fundamental frequencies:
\begin{equation}
\omega_{mnk} = m\Omega_\phi + n\Omega_r + k\Omega_\theta.
\label{eq:mode_freq}
\end{equation}
For a snapshot at a large luminosity distance $D_L$, the waveform in the source-centered frame with the polar axis aligned with the spin of the central black hole is given by:
\begin{equation}
\begin{split}
h(t) = \frac{\mu}{D_L} \sum_{\ell,m,k,n} & A_{\ell m n k} \, {}_{-2}S_{\ell m}( a\omega_{mnk}; \theta_S) \\
& \times e^{-i\omega_{mnk} t + i m \phi_S}.
\end{split}
\label{eq:waveform_snapshot}
\end{equation}
The angles $(\theta_S, \phi_S)$ specify the direction of the observer in this frame, and ${}_{-2}S_{\ell m}$ are the spin-weighted spheroidal harmonics. In this work, the surrogate is trained directly on amplitudes in this spheroidal basis. The complex coefficients $A_{\ell m n k}$ modeled here are the snapshot amplitudes for a fiducial geodesic. Amplitudes for the same orbit with arbitrary initial orbital phases can be recovered by a known phase correction~\cite{Drasco:2005is, Hughes:2021exa, Burke:2023lno}.

The complexity of the mode spectrum is directly tied to the orbital geometry defined in Sec.~\ref{sec:orbit_geometry}. For instance, circular orbits ($e=0$) have no radial motion, suppressing all modes with $n \neq 0$. Similarly, equatorial orbits ($|x_I|=1$) have no polar motion, suppressing modes with $k \neq 0$. A generic Kerr (KG) orbit, however, excites the full spectrum of modes. The sheer number of contributing harmonics—often on the order of ${\sim 10^5}$ or more for a single snapshot~\cite{Drasco:2005kz, Hughes:2021exa}—coupled with the need to evaluate their amplitudes $A_{\ell m n k}$ across the four-dimensional orbital parameter space $(a, p, e, x_I)$, gives rise to the well-known ``amplitude bottleneck'' in EMRI modeling.

Our work directly targets this bottleneck by creating a surrogate for the complete set of snapshot amplitudes, $A_{\ell m n k}$.
The construction of a full inspiral waveform then involves evaluating these surrogate-generated amplitudes along a pre-computed radiation-reaction trajectory, a procedure known as adiabatic evolution~\cite{Hinderer:2008dm}.
This approach is well-posed for two key reasons. First, while accurate phase evolution ultimately requires \ac{PA} corrections, the amplitudes themselves are only needed at this leading adiabatic order for EMRI data analysis~\cite{Hinderer:2008dm, Burke:2023lno}.
Second, the amplitudes for any initial orbital phase can be recovered from those of a single fiducial geodesic via a simple phase correction, meaning the surrogate does not need to learn this extra degree of freedom~\cite{Drasco:2005is, Hughes:2021exa}.

We therefore focus on predicting the fiducial snapshot amplitudes.
To ensure comprehensive coverage of the spectrum, our model targets modes spanning $\ell \in [2, 10]$, $n \in [-50, 50]$, and $k \in [-10, 10]$.
Exploiting a well-known conjugate symmetry for the amplitudes of negative-$m$ modes~\cite{Hughes:1999bq, Fujita:2020zxe}, we only model the $m \ge 0$ sector.
This defines a target space of 133,614 unique complex amplitudes per snapshot, which our surrogate aims to predict as a single function of the orbital parameters $(a, p, e, x_I)$.

\subsection{Snapshot Amplitudes from Teukolsky formalism}
The snapshot amplitudes are derived from solutions to the Teukolsky equation~\cite{Teukolsky:1973ha}, which governs the dynamics of a gauge-invariant Weyl curvature scalar, $\psi_4$, in a perturbed Kerr spacetime.
For a point-particle source of mass $\mu$ on a geodesic, the equation is linear and separable.
By Fourier decomposing $\psi_4$ in time and expanding in spin-weighted spheroidal harmonics ${}_{-2}S_{\ell m}(\theta; a\omega)$ for the angular dependence, the problem reduces to solving an ordinary differential equation for the radial function $R_{\ell m \omega}(r)$ for each frequency $\omega$ and multipole $(\ell, m)$:
\begin{equation}
\Delta^2 \frac{d}{dr}\left(\frac{1}{\Delta}\frac{dR_{\ell m \omega}}{dr}\right) - V(r) R_{\ell m \omega}(r) = \mathcal{T}_{\ell m \omega}.
\end{equation}
Here, $\Delta = r^2 - 2Mr + a^2$, $V(r)$ is a complex potential that depends on the frequency and black hole parameters, and $\mathcal{T}_{\ell m \omega}$ is the source term derived from the particle's trajectory.

The solution to this radial equation must satisfy physical boundary conditions: purely outgoing waves at infinity ($r \to \infty$) and purely ingoing waves at the event horizon ($r \to r_+$).
This involves constructing the full solution by integrating the source term against a Green's function built from homogeneous solutions that satisfy these boundary conditions~\cite{Poisson:2011nh}.
The asymptotic behavior of the radial solution yields the complex amplitudes at infinity, $Z^\infty_{\ell m n k}$, and at the horizon, $Z^H_{\ell m n k}$, for each discrete mode $(\ell, m, n, k)$ present in the source's spectrum~\cite{Teukolsky:1973ha, Hughes:1999bq}.
The waveform amplitudes are directly proportional to the amplitudes at infinity, $A_{\ell m n k} \propto Z^\infty_{\ell m n k} / \omega_{mnk}^2$, while the gravitational-wave fluxes needed to compute the inspiral are constructed from the sum over modes of $|Z^\infty_{\ell m n k}|^2$ and $|Z^H_{\ell m n k}|^2$~\cite{Teukolsky:1973ha, Hughes:1999bq}.
This procedure, which must be repeated for each of the ${\sim 10^5}$ modes at every point in the 4D parameter space, represents the primary computational obstacle to rapid waveform generation.

Two principal methodologies exist for solving the Teukolsky equation and obtaining these amplitudes.
Numerical solvers, such as those provided by the Black Hole Perturbation Toolkit (BHPT)~\cite{BHPToolkit}, offer high-accuracy solutions valid throughout the strong-field regime.
However, their substantial computational cost makes them impractical for generating the large, densely-sampled datasets required for surrogate model training across the full 4D parameter space.
Conversely, semi-analytical methods, primarily based on Post-Newtonian (PN) expansions, provide computationally efficient analytical expressions for the amplitudes~\cite{Mino:1997bx, Sasaki:2003xr}.
These models expand the solution in powers of orbital velocity and eccentricity, offering rapid evaluation at the expense of being formally valid only in the weak-field, low-eccentricity limit.
To develop and validate our surrogate modeling framework across the full dimensionality of the generic Kerr problem, this work uses a semi-analytical dataset, which provides a practical, low-cost testbed for assessing the scalability of our architecture.

\section{Methodology} \label{sec:methodology}
Our core strategy is to replace the direct, computationally intensive snapshot calculation of Teukolsky amplitudes with a fast and accurate surrogate model.
This section details the proposed framework, which is designed to be general and extensible.
We describe the model architecture, the training methodology, and briefly introduce the training and validation sets.

\begin{figure}
    \centering
    \includegraphics[width=1\linewidth]{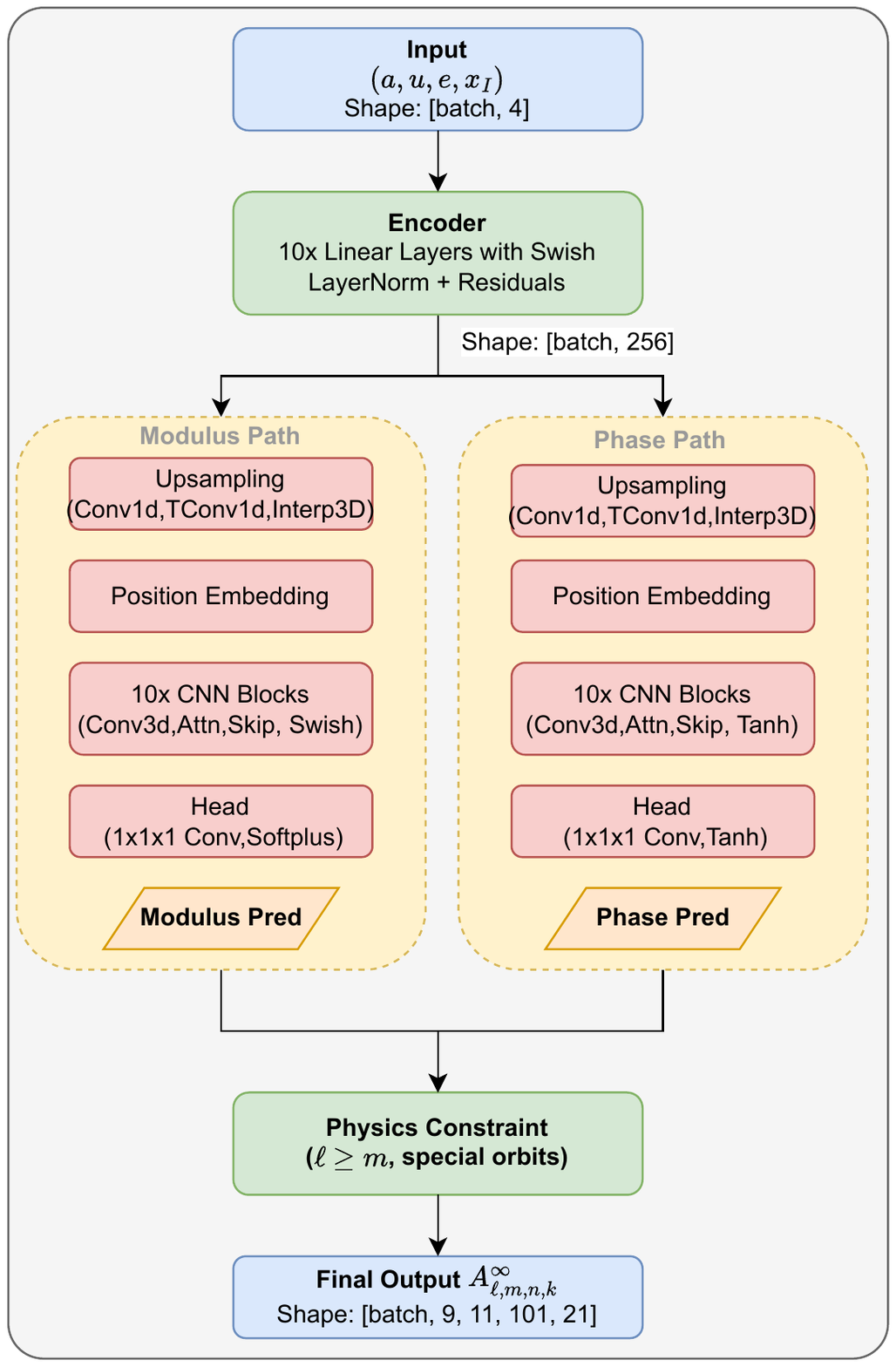}
    \caption{The neural network employs an encoder-decoder architecture to predict the Teukolsky amplitude's modulus and phase in parallel branches. The encoder, a 10-layer residual MLP with Swish activations, maps the four orbital parameters $(a, p, e, x_I)$ to a latent vector. This vector is then upsampled using trilinear interpolation along  $(m,n,k)$, with the $\ell$ index represented by the channel dimension. A series of 10 residual CNN blocks, featuring 3D convolutions with anisotropic kernels and attention gates, refines the structural tensor. Independent output heads with physically-motivated activations (Softplus for the modulus and Tanh for the phase) produce the predictions, which are passed through a final layer that enforces physical constraints (e.g., $|m| \le \ell$).}
    \label{fig:architecture_diagram}
\end{figure}

\subsection{Model Architecture}
Our surrogate is built upon a convolutional encoder-decoder architecture, as illustrated in Fig.~\ref{fig:architecture_diagram}, designed specifically for the task of mode prediction. The architecture consists of four main components: (1) an encoder that maps the 4D physical parameter space $(a, p, e, x_I)$ to a compact latent representation; (2) a convolutional decoder that synthesizes the full, high-dimensional amplitude tensor from this representation; (3) parallel output heads that produce the final modulus and phase predictions; and (4) a final layer that enforces physical constraints.

\textbf{Encoder.} The encoder path begins by processing the input orbital parameters $\mathbf{v} = (a, p, e, x_I)$. To handle the diverse physical scales of these parameters, they are first individually normalized to the range $[-1, 1]$ using a min-max scaling based on their predefined physical boundaries. The normalized vector $\mathbf{v'}$ is then projected into a higher-dimensional feature space using a trainable Fourier feature mapping layer~\cite{tancik2020fourier}. This layer, defined by the function $\gamma(\mathbf{v'}) = [\cos(2\pi \mathbf{Bv'}), \sin(2\pi \mathbf{Bv'})]^T$ with a trainable frequency matrix $\mathbf{B} \in \mathbb{R}^{128 \times 4}$, transforms the 4-dimensional input into a 256-dimensional feature vector. This projection enables the network to adaptively learn the optimal basis functions required to capture complex, high-frequency dependencies between the orbital parameters and the amplitude output~\cite{rahaman2019spectral}. The resulting feature vector is then processed by a deep Multi-Layer Perceptron (MLP) consisting of 10 residual blocks. Each block comprises a Linear layer, Layer Normalization~\cite{ba2016layer}, and a Swish activation function, distilling the features into a final latent vector $\mathbf{z} \in \mathbb{R}^{256}$.

\textbf{Decoder.} The decoder synthesizes the final amplitude tensor from the latent vector $\mathbf{z}$ in a two-stage process. First, in an \textit{initial upsampling} stage, $\mathbf{z}$ is transformed into a small, channel-rich intermediate tensor $\mathcal{T}_0 \in \mathbb{R}^{N_\ell \times D_m \times D_n \times D_k}$ through a learnable sequence of 1D transposed convolutions. This small tensor is then efficiently upsampled to the full target mode-space dimensions $(N_m, N_n, N_k) = (11, 101, 21)$ via trilinear interpolation, which establishes a coarse global structure for the amplitude tensor. To make subsequent operations position-aware and break the translation-equivariance of convolutions, the upsampled tensor is concatenated with a sinusoidal positional embedding that explicitly encodes the normalized coordinates of each point in the $(m,n,k)$ mode grid.

Second, in the \textit{refinement} stage, a series of 10 residual RefinementBlocks iteratively improves this structured tensor. Each block is built upon a 3D convolution and contains Group Normalization~\cite{wu2018group} and a Swish activation. Crucially, these blocks employ two advanced features: (i) 3D convolutions with \textit{anisotropic kernels and dilations} (e.g., with shapes like $(9, 3, 1)$ and $(1, 27, 1)$), which create receptive fields tailored to the different physical correlation lengths along the $m$, $n$, and $k$ axes; and (ii) a 3D \textit{attention gate}, inspired by Attention U-Net~\cite{oktay2018attention}, which allows the network to dynamically re-weight features and focus on the most salient regions of the mode space during refinement.

\textbf{Output Heads \& Physics Constraints.} The single refined tensor from the decoder is fed into two independent, parallel output heads for the final predictions. Each head consists of a simple 1x1x1 Conv3d layer, which acts as a per-pixel linear projection, followed by a physically-motivated activation function: `Softplus` for the non-negative modulus and the hyperbolic tangent (tanh) for the bounded phase. Finally, a non-trainable layer applies a binary mask to the output, enforcing known physical selection rules. This mask sets to zero all mode amplitudes that are known to vanish, such as those violating the spherical-harmonic index rule ($|m| > \ell$), the static mode ($m=k=n=0$), and modes suppressed by orbital symmetries (e.g., $n\neq0$ for circular orbits and $k\neq0$ for equatorial orbits), ensuring the surrogate's output is always physically valid.

\subsection{Training Dataset and Preprocessing}
The foundational dataset for this work is provided by the \ac{BHPC}~\cite{Isoyama:2021jjd}, based on a semi-analytical solution of the Teukolsky equation.
The data are calculated to 5PN order in velocity ($v=\sqrt{1/p}$) and 10th order in eccentricity (5PN-$e^{10}$), providing a computationally inexpensive means to generate amplitudes across the full parameter space of generic orbits.
This allows us to rigorously test the scalability of our framework, a task that remains prohibitive with purely numerical solvers.
The dataset provides amplitudes for approximately 33,000 unique harmonic modes, covering indices up to $2 \le \ell \le 12$, $|m| \le \ell$, $|m+k| \le 12$, and $|n| \le 10$.

The full dataset contains 25880 samples, including $12940$ grid-based samples and $12940$ random samples.
Specifically, the training set is composed of all grid-based samples plus 80\% of the random samples, while the validation set consists of the remaining 20\% of the random samples.
Since the amplitudes vary more rapidly in the strong-field regime near the LSO,
the sample density should be higher as the semi-latus rectum $p$ becomes smaller.
So, for both types of sampling, we adopt the coordinate transformation
$u= 1/\sqrt{p - 0.9 p_{\text{LSO}}}$,
where $p_{\text{LSO}}$ is the semi-latus rectum of the last stable orbit, which is itself a function of $(a, e, x_I)$~\cite{Stein:2019buj}.
Then, if we have a uniform random sampling on $u$, we will have a higher density for small $p$.

\begin{figure}[!htbp]
  \centering
  \includegraphics[width=\linewidth]{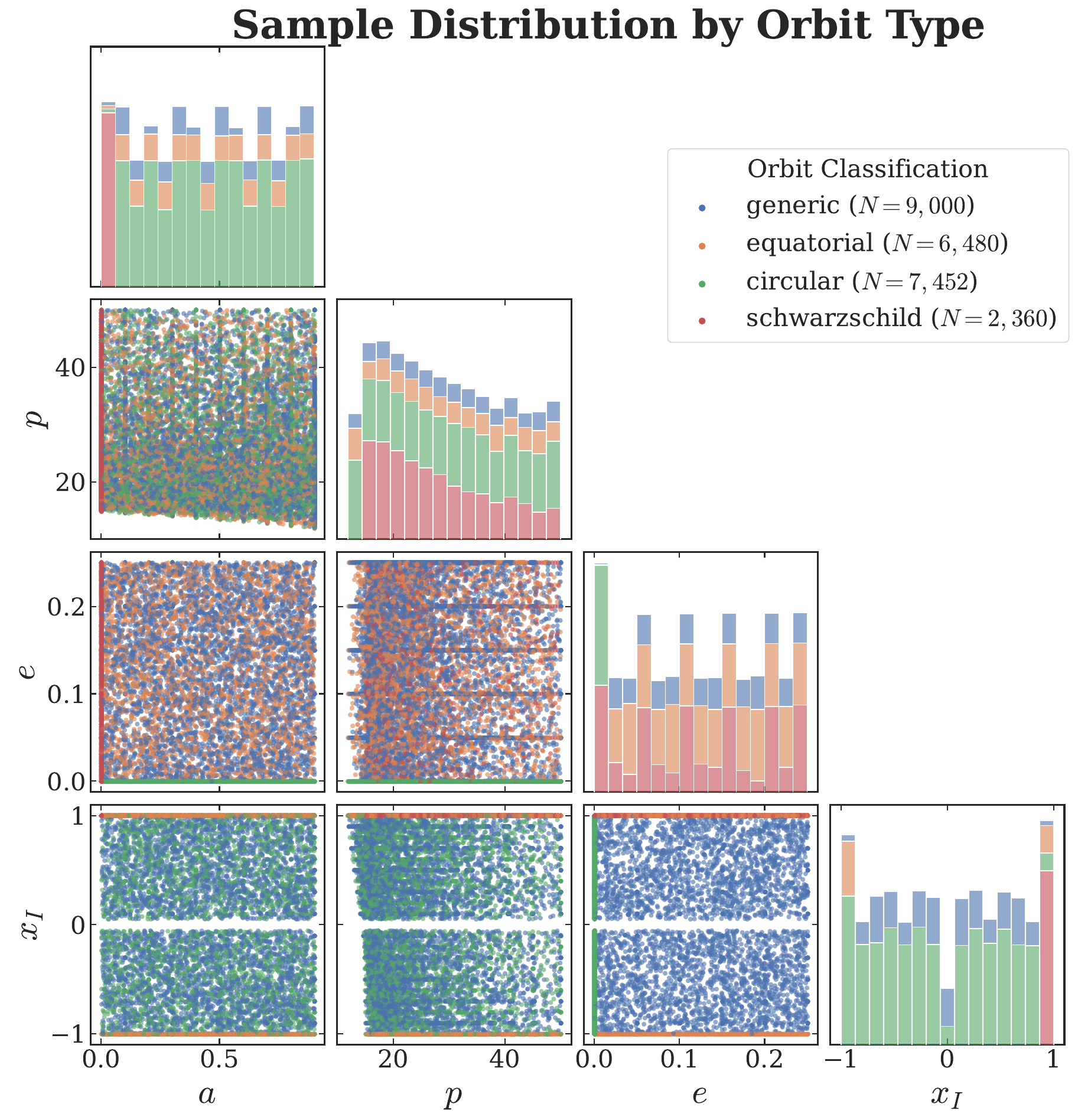}
   \caption{Distribution of orbital parameters in the \textbf{training dataset}.
The corner plot shows 1D marginalized histograms (diagonal) and 2D projected distributions (off-diagonal) for the spin $a$, semi-latus rectum $p$, eccentricity $e$, and inclination cosine $x_I$.
The color map distinguishes the different orbital geometries as defined in Table~\ref{tab:orbit_classification}.
The plot visualizes the dataset's stratified nature, with dense populations corresponding to specific classes like Schwarzschild (SC/SE), Kerr Equatorial (KEC/KEE), Kerr Inclined Circular (KIC), and Kerr Generic (KG) orbits.}
   \label{fig:train_dist}
\end{figure}

\begin{figure}[!htbp]
   \centering
   \includegraphics[width=\linewidth]{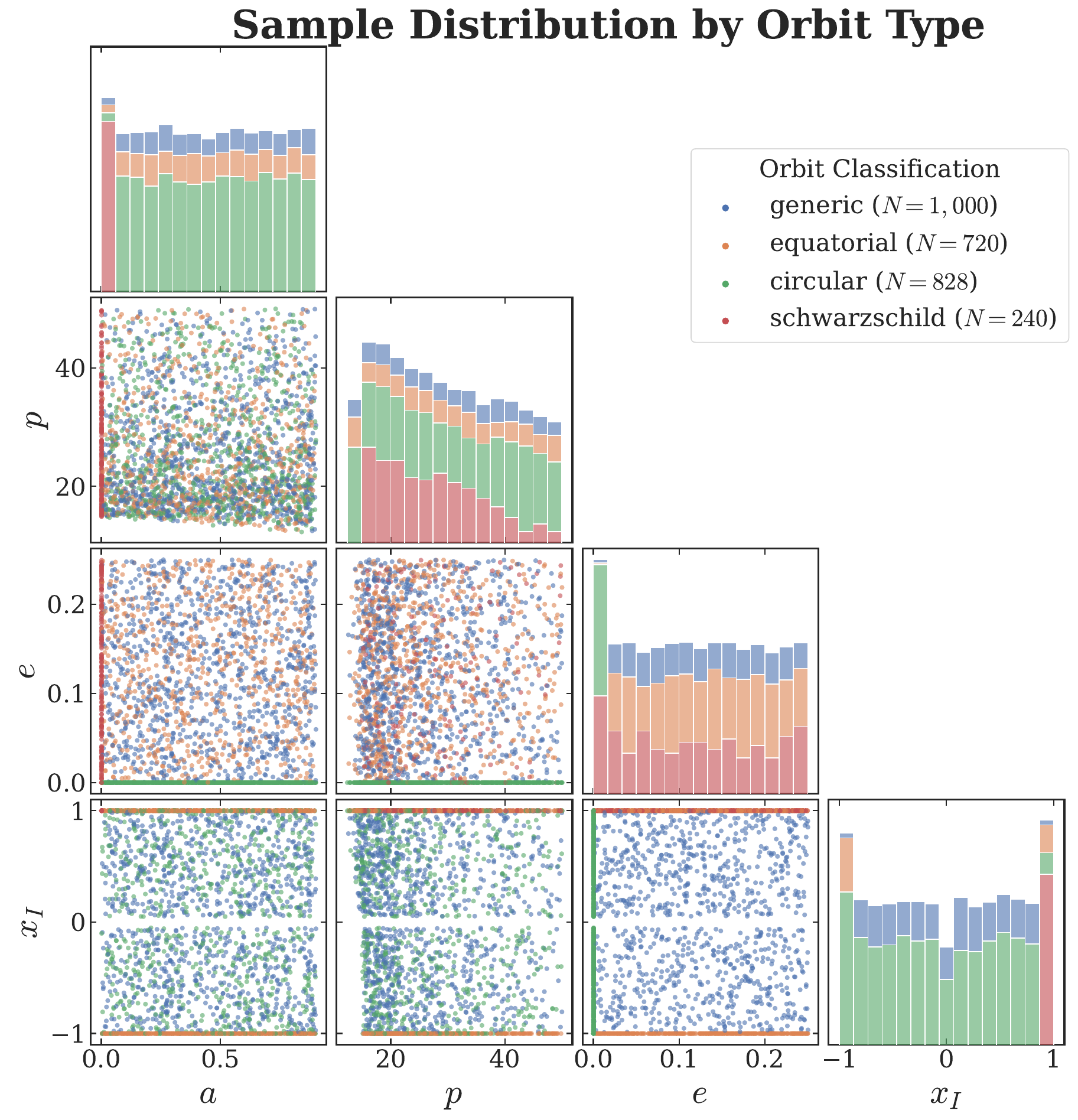}
   \caption{Distribution of orbital parameters in the \textbf{validation dataset}.
This dataset consists of randomly drawn samples that were held out from the training process.
It covers all orbital classes, providing a robust test of the model's ability to generalize to unseen data.}
   \label{fig:valid_dist}
\end{figure}

The samples are generated separately for different types of orbits, and the details are listed below.
 \begin{enumerate}[leftmargin=*,label=\textbf{Case \arabic*}:]
    \item \textbf{SC Grid} (200 samples) \\
    $a=0,~x_I=1,~e=0$. \\
    $p$: 200 equidistant samples on $u$.

    \item \textbf{SE Grid} (1000 samples) \\
    $a=0,~x_I=1$. \\
    $e \in \{0.05, 0.1, 0.15, 0.2, 0.25\}$.\\
    $p$: 200 equidistant samples on $u$.

    \item \textbf{KEC Grid} (900 samples) \\
    $a \in \{0.1, 0.2, \dots, 0.9\}$. \\
    $|x_I|=1,~e=0$. \\
    $p$: 50 equidistant samples on $u$.

    \item \textbf{KEE Grid} (3600 samples) \\
    $a \in \{0.1, 0.2, \dots, 0.9\}$. \\
    $|x_I|=1$. \\
    $e \in \{0.05, 0.1, 0.15, 0.2, 0.25\}$.\\
    $p$: 40 equidistant samples on $u$.

    \item \textbf{KIC Grid} (3240 samples) \\
    $a \in \{0.1, 0.2, \dots, 0.9\}$. \\
    $x_I \in \{-0.9, -0.8, \dots, -0.1, 0.1, \dots, 0.8, 0.9\}$.\\
    $e=0$. \\
    $p$: 20 equidistant samples on $u$.

    \item \textbf{KG Grid} (4000 samples) \\
    $a \in \{0.1, 0.3, 0.5, 0.7, 0.9\}$. \\
    $x_I \in \{-0.9, -0.6, -0.3, -0.1, 0.1, 0.3, 0.6, 0.9\}$.\\
    $e \in \{0.05, 0.1, 0.15, 0.2, 0.25\}$. \\
    $p$: 20 equidistant samples on $u$. \\
\end{enumerate}
Here, $p\in[p_{\text{min}},50]$, and $p_{\text{min}}=\max(p_{\text{LSO}},10)$.
So the samples in $p$ are determined after we choose a specific combination of $a,~x_I,~e$.
The lower cutoff at $p=10$ and $e=0.25$ is a conservative choice
motivated by the behavior of the PN dataset itself:
for smaller-$p$ and larger-$e$ regions, 
some amplitudes developed very large excursions, 
indicating that the PN expansion was poorly behaved in these regions. 
We therefore restrict the present validation study to a numerically better-behaved region so that the neural-network framework can be assessed on a cleaner dataset.
For each orbit type, the random samples use the same parameter ranges and the same number of samples as the corresponding grid samples. 
The distributions for the training and validation sets are shown in the corner plots in Fig.~\ref{fig:train_dist} and Fig.~\ref{fig:valid_dist}. The distributions of spin $a$, eccentricity $e$ and inclination cosine $x_I$ show distinct peaks at their boundary values. This corresponds to different orbital geometries. For instance, Schwarzschild cases are confined to the $a=0$ line, while peaks at the boundaries correspond to specific geometries, such as equatorial orbits at $|x_I|=1$ and circular orbits at $e=0$. 
A small neighborhood of the polar limit, $|x_I|\le 0.05$, is intentionally excluded from both datasets. Strictly polar orbits are technically difficult for Teukolsky-amplitude calculations, and in practice this exclusion helps keep the training set numerically well behaved. This makes the parameter domain non-continuous across the polar region.

Several necessary preprocessing steps were applied to the raw dataset to ensure its suitability for training.
First, we address the issue of orbital resonances.
For certain orbital configurations, the fundamental frequencies satisfy a commensurability condition, $m\Omega_\phi + k\Omega_\theta + n\Omega_r \approx 0$, causing the mode frequency $\omega_{mnk}$ to approach zero.
In the snapshot formalism, this leads to a numerical divergence in the waveform amplitude $A_{\ell m n k} \propto \omega_{mnk}^{-2}$, which destabilizes the training process.
A rigorous treatment of resonances requires going beyond the adiabatic approximation~\cite{Hughes:2016xwf, Flanagan:2010cd, Isoyama:2021jjd, LISAConsortiumWaveformWorkingGroup:2023arg}.
Accordingly, we identify and exclude resonant samples from our training set by filtering for points where $|\omega_{mnk}| < 10^{-3}$ while $|A_{\ell m n k}| > 1$.
This removes approximately 1\% of the generic orbit data, allowing us to focus on fitting the non-resonant amplitude function.
Second, we must reconcile differing orbital conventions.
As stated in Sec.~\ref{sec:orbit_geometry}, our framework parameterizes inclination via $x_I \in [-1, 1]$.
The \ac{BHPC} dataset, however, uses a different convention for retrograde orbits: its amplitudes $A'_{\ell mnk}$ are tabulated with an inclination variable $Y \in [0,1]$, while retrograde configurations are represented by a negative spin parameter. In our convention, $x_I \in [-1,1]$ directly distinguishes prograde and retrograde orbits. The mapping used in our preprocessing is
\begin{equation}
\begin{gathered}
A_{\ell mnk}(a,p,e,x_I)= \\[-0.3ex]
{}
\begin{cases}
\begin{aligned}[t]
&A'_{\ell mnk}(q=a,p,e,Y=x_I),
\end{aligned}
& x_I\ge 0, \\[1ex]
\begin{aligned}[t]
&(-1)^\ell A'_{\ell,-m,n,k}(q=-a,p,e,Y=-x_I),
\end{aligned}
& x_I<0 .
\end{cases}
\end{gathered}
\end{equation}

To create a consistent tensor structure for our model, we must therefore pad the amplitude array with zeros for modes that are undefined under a given convention.
This zero-padding is used to place all samples on a common fixed output grid. A small neighborhood around the polar limit, $|x_I|\le 0.05$, is excluded from the dataset to keep the samples numerically well behaved.

\subsection{Training Strategy}
The core of our approach is a curriculum-based transfer learning strategy.
The model is trained in a sequence of stages, starting with the simplest physical scenarios and progressively incorporating more complex orbital configurations, as defined in Table~\ref{tab:orbit_classification}.
At each stage, the dataset is augmented with samples from the new orbital class, and the model obtained from the previous stage serves as the weight initialization for the new training run. 
The stages proceed in the order of increasing complexity:
SC/SE $\to$ KEC $\to$ KEE $\to$ KIC $\to$ KG.

This curriculum learning strategy is motivated by several factors.
First, it allows the model to master fundamental dependencies in simpler, lower-dimensional spacetimes (e.g., on $p$ and $e$ in Schwarzschild) before incorporating the effects of spin $a$ and inclination $x_I$.
Second, this staged progression provides a more stable and effective training path than presenting the model with a highly diverse dataset from scratch. This is particularly relevant for future applications where data for generic orbits will be computationally expensive; by transferring knowledge from simpler, cheaper-to-generate orbital configurations, we hypothesize that the data requirements for the most complex classes can be reduced.

The network is trained by minimizing a composite loss function,
\begin{equation}
L = w_{\text{rel}} L_{\text{rel}} + w_{\text{cross}} L_{\text{cross}} + w_{\phi} L_{\phi}.
\end{equation}
This loss function is a weighted sum of three distinct components, each designed to enforce a different aspect of prediction.
We found that the weights $(w_{\text{rel}}, w_{\text{cross}}, w_{\phi}) = (0.8, 0.1, 0.1)$ provide a balance between fitting the individual mode amplitudes, their overall distribution, and their phases.
The components are as follows:

\begin{enumerate}[leftmargin=*]
    \item \textbf{Relative Amplitude Loss ($L_{\mathrm{rel}}$):} This component targets the accuracy of individual mode amplitudes. A standard Mean Absolute Error (MAE) struggles with the vast dynamic range of the overall amplitude scale, which can vary by orders of magnitude across the orbital parameter space. To ensure uniform fitting fidelity, we define a self-normalizing relative error. For each sample, the MAE is scaled by the magnitude of that sample's single strongest true mode, $|\mathcal{A}_\text{top1}^{\text{true}}|$:
    \begin{equation}
        L_{\mathrm{rel}}=\sum_i\frac{\left|\mathcal{A}_i^{\mathrm{pred}}-\mathcal{A}_i^{\text {true }}\right|}{\left|\mathcal{A}_\text {top1}^{\text {true }}\right|}.
    \end{equation}
    $\{\mathcal{A}_i\}$ denotes a vectorization of $\{A_{\ell m n k}\}$ over the chosen mode set.
    This formulation preserves the absolute error's sensitivity to discrepancies in dominant modes while normalizing the loss across samples with vastly different energy scales, thereby preventing the training process from being dominated by high-amplitude samples and promoting a more uniform fitting across the entire parameter space.

    \item \textbf{Distributional Cross-Entropy Loss ($L_{\text{cross}}$):} While $L_{\text{rel}}$ focuses on the precise fit of individual (and often dominant) modes, it can be insensitive to the collective distribution of the many weaker modes. To address this, we introduce a loss component from the domain of classification: cross-entropy. We treat the normalized magnitude spectrum as a probability distribution, compelling the model to learn the overall shape and energy allocation across all modes. The sets of predicted and true amplitude magnitudes are normalized to sum to unity, forming discrete probability distributions $Q$ and $P$, respectively. The loss is their cross-entropy:
    \begin{equation}
        L_{\text{cross}} = - \sum_i P_i \log Q_i,
    \end{equation}
    where $P_i = |\mathcal{A}_i^{\text{true}}| / \sum_j |\mathcal{A}_j^{\text{true}}|$ and $Q_i = |\mathcal{A}_i^{\text{pred}}| / \sum_j |\mathcal{A}_j^{\text{pred}}|$. This macroscopic constraint guides the model to fit the amplitudes at the right order of magnitude and is highly sensitive to the overall distributional shape. It ensures the surrogate correctly reproduces the collective behavior of sub-dominant modes, which is important for capturing the full morphology of the waveform, rather than just fitting the few most powerful modes.

    \item \textbf{Weighted Phase Loss ($L_{\phi}$):} The final waveform is constructed from the coherent sum of all modes, making phase accuracy critical. However, the impact of phase errors is not uniform; errors in high-amplitude modes are far more detrimental than those in weak modes. We therefore introduce a weighted phase loss that prioritizes the phase accuracy of the most energetically significant modes. It measures the shortest angle difference between the predicted and true phase, weighted by the mode's relative amplitude:
    \begin{equation}
        L_\phi=\frac{1}{N} \sum_i w_i \cdot|\operatorname{atan2}\left(\sin \left(\Delta \phi_i\right), \cos \left(\Delta \phi_i\right)\right)|,
    \end{equation}
    where $\Delta \phi_i = \phi_i^{\text{pred}} - \phi_i^{\text{true}}$ and the weight $w_i = |\mathcal{A}_i^\text{true}|/|\mathcal{A}_\text{top1}^\text{true}|$. The use of atan2 correctly handles the $2\pi$ periodicity of the phase. This weighting scheme ensures that the model dedicates its capacity to precisely learning the phases of the modes that have the greatest impact on the final waveform's fidelity.
\end{enumerate}
In preliminary experiments, we observed that training with only a relative amplitude loss ($L_{\text{rel}}$) resulted in models that, while accurately predicting the dominant modes, failed to capture the correct power distribution across the broader spectrum of weaker modes.
Adding the cross-entropy term ($L_{\text{cross}}$) corrected this distributional shape,
and the weighted phase loss ($L_{\phi}$) reduced the final waveform mismatch significantly, as unweighted phase errors in sub-dominant modes led to significant dephasing in the coherently summed signal.
In our training, the cross-entropy term mainly acts as an auxiliary constraint on the global spectral structure, especially in the earlier stage of optimization.

The network is trained by minimizing the composite loss function described above using the AdamW optimizer~\cite{loshchilov2017decoupled}.
We employ a batch size of 1280 and set the initial learning rate to $3 \times 10^{-4}$.
The learning rate schedule begins with a linear warm-up over the first 400 steps to ensure initial stability.
Following the warm-up, a ReduceLROnPlateau scheduler is used, which reduces the learning rate by a factor of 0.9 if the validation loss does not improve for 150 epochs~\cite{paszke2019pytorch}.
For each stage of our curriculum learning, the model is trained for a fixed duration of 20,000 epochs, providing ample time for convergence.
A notable aspect of our training is the use of a strong weight decay ($\lambda=1.0$). We found this high level of regularization to be effective for preventing overfitting, particularly during the early stages of the curriculum on simpler, lower-dimensional datasets, and for ensuring that the learned features remain general enough for effective transfer to more complex orbital geometries.
The entire training process is performed in 32-bit precision on a distributed setup of 4x NVIDIA A100 GPUs.
To accommodate the large memory footprint of our model architecture, we utilize gradient checkpointing~\cite{chen2016training}, which trades computational time for reduced memory usage by re-computing intermediate activations during the backward pass.

\section{Results} \label{sec:result}
\begin{figure}[!htbp]
   \centering
   \includegraphics[width=\linewidth]{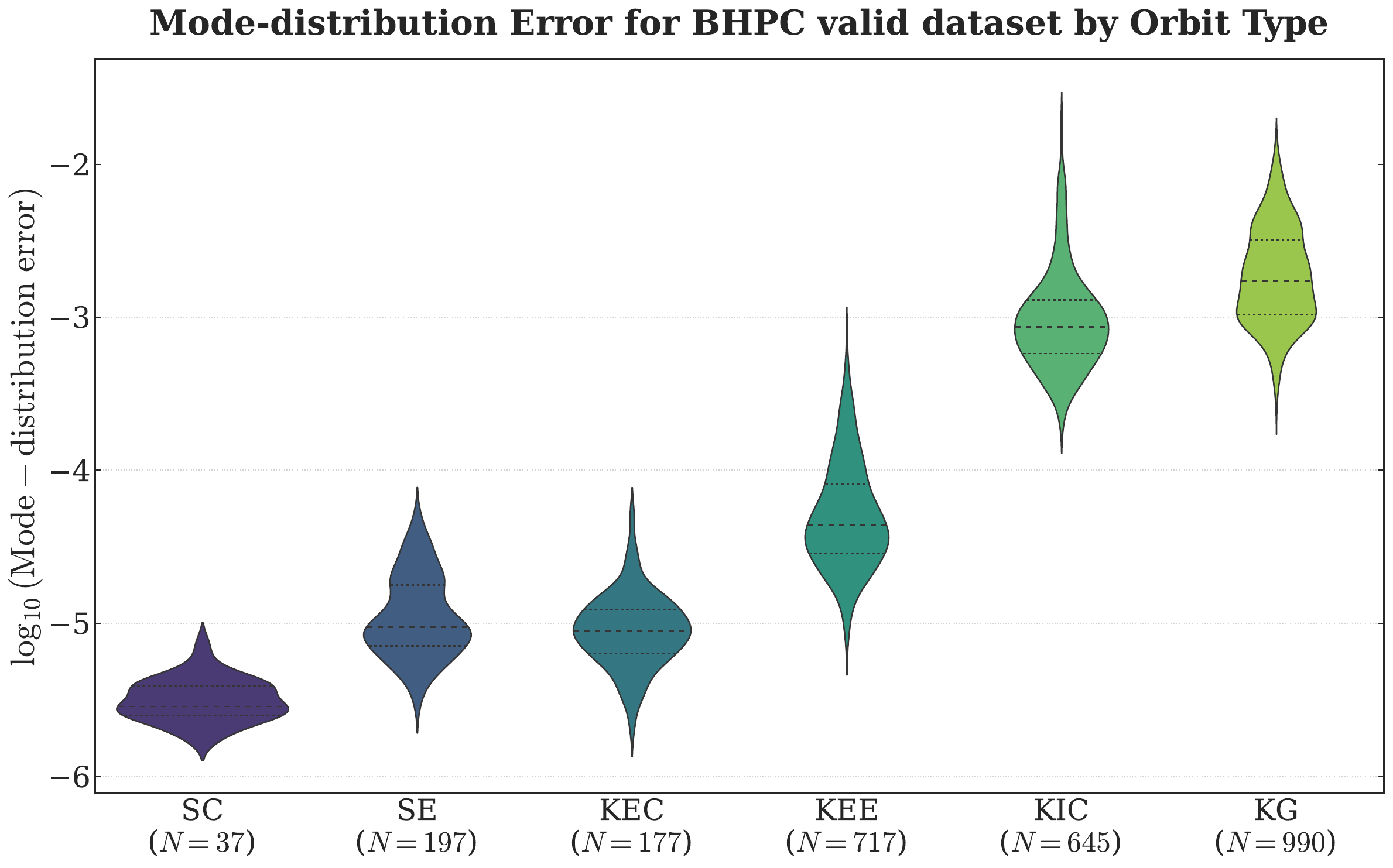}
   \caption{The mode-distribution error ($\mathcal{M}_\text{amp}$) categorized according to orbital geometry on a logarithmic scale. The shape of each violin shows the probability density of the error, while the inner box plot marks the median and interquartile range.}
   \label{fig:mode_distribution_error}
\end{figure}

The performance of our surrogate model is evaluated on the PN validation dataset. 
Our primary metric for accuracy is the mode-distribution error, $\mathcal{M}_\text{amp}$, which measures the mismatch between the predicted and true complex amplitude vectors:
\begin{equation}
    \mathcal{M}_\text{amp} = 1 - \frac{\Re\left(\sum_{i} \mathcal{A}_{i}^{\text{pred}} \cdot (\mathcal{A}_{i}^{\text{true}})^*\right)}{\left|\mathcal{A}^{\text{pred}}\right| \cdot \left|\mathcal{A}^{\text{true}}\right|}.
    \label{eq:mode-distribution-error}
\end{equation}
While high-precision parameter estimation ultimately demands even smaller waveform mismatches, a mode-distribution error of $\mathcal{M}_\text{amp} \lesssim 10^{-2}$ is a widely accepted baseline requirement to ensure unbiased signal extraction for a significant fraction of sources~\cite{Isoyama:2021jjd, Chapman-Bird:2025xtd}.

Figure~\ref{fig:mode_distribution_error} shows the distribution of $\mathcal{M}_\text{amp}$ across the different orbital geometries.
A clear trend is visible: the model's accuracy degrades as orbital complexity increases.
For simpler geometries—Schwarzschild Circular (SC), Schwarzschild Eccentric (SE), and Kerr Equatorial Circular (KEC)—the median error is low, clustering between $10^{-5}$ and $10^{-6}$.
The error grows for more complex cases, with the median increasing to approximately $3 \times 10^{-5}$ for Kerr Equatorial Eccentric (KEE) orbits, and further to around $10^{-3}$ for Kerr Inclined Circular (KIC) and Kerr Generic (KG) orbits.
The widening of the error distributions for these latter cases also indicates larger performance variance.
This trend is an expected consequence of the ``curse of dimensionality.'' The parameter space grows from two dimensions for Schwarzschild eccentric systems to four for generic Kerr orbits, while the output space of harmonic modes becomes vastly richer.
For a fixed number of training samples, the effective sample density in this high-dimensional space becomes increasingly sparse, making it inherently more challenging for the model to learn the intricate dependencies required for a globally accurate fit.

\begin{table}[h!]
\small
\caption{Recall scores for the identification of dominant modes (those contributing to 99\% of total power). Performance is high for simpler orbits but degrades for inclined and generic configurations.}
\label{tab:recall_scores}
\begin{ruledtabular}
\begin{tabular}{lc}
\textbf{Orbit Type} & \textbf{Recall Score (\%)} \\
\hline
Schwarzschild Circular (SC) & 100.0 \\
Schwarzschild Eccentric (SE) & 99.7 \\
Kerr Equatorial Circular (KEC) & 99.7 \\
Kerr Equatorial Eccentric (KEE) & 99.0 \\
Kerr Inclined Circular (KIC) & 94.0 \\
Kerr Generic (KG) & 83.9 \\
\end{tabular}
\end{ruledtabular}
\end{table}

Beyond the overall error, we also evaluate the model's ability to correctly identify the set of dominant modes, which carry the vast majority of the radiated power.
Following Drasco and Hughes~\cite{Drasco:2007gn}, who showed that a few hundred modes typically account for 99\% of the power, we operationally define the ``dominant mode set'' as the smallest set of modes whose cumulative power accounts for 99\% of the total, with each mode individually contributing at least 0.1\%.
Table~\ref{tab:recall_scores} shows the recall score for this set across orbital types.
The model achieves near-perfect recall (99.7-100\%) for Schwarzschild and Kerr equatorial orbits. However, the performance drops for inclined systems, with the recall for KIC orbits at 94.0\% and degrading to 83.9\% for fully generic (KG) orbits, which can have $\sim 10^5$ excited modes. This indicates that while the model captures the overall power distribution, it struggles to precisely identify every significant mode in the most complex scenarios.

\begin{figure}[!htbp]
   \centering
   \includegraphics[width=\linewidth]{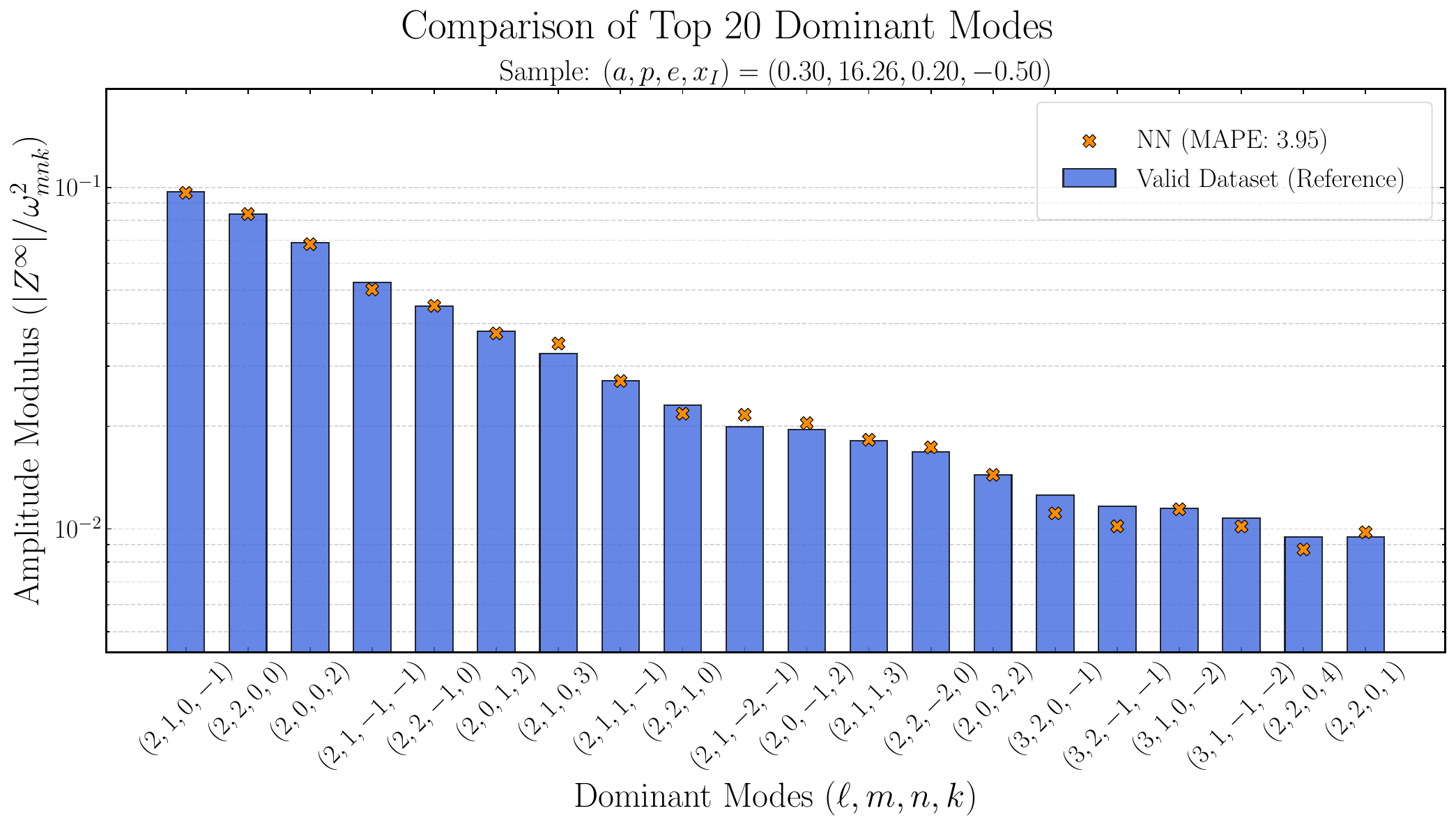}
   \caption{Comparison of the predicted and true log-magnitudes for the top 20 dominant modes of a representative Kerr Generic (KG) orbit.
The specific orbital parameters are $(a, p, e, x_I) = (0.30, 16.26, 0.20, -0.50)$.
The x-axis lists the mode indices $(\ell, m, n, k)$ for each of the 20 modes, ordered by their true amplitude.
The y-axis shows the amplitude modulus, $|A_{\ell m n k}|$, on a logarithmic scale.
Blue bars represent the true amplitudes from the PN dataset (Reference), while orange bars show the corresponding predictions from our surrogate model (NN).
The Mean Absolute Percentage Error (MAPE) for this specific sample is 3.95\%.}
   \label{fig:top20_comparison}
\end{figure}

$\mathcal{M}_\text{amp}$ characterizes the quantity $1 - \cos\theta$ (the cosine of the angle between two complex-valued vectors), rendering it invariant to global scaling of the magnitudes.
Consequently, if the predicted values differ from the ground truth by a constant multiplicative factor, $\mathcal{M}_\text{amp}$ remains vanishingly small, failing to capture absolute amplitude discrepancies.
To visualize the performance of the model in a challenging case, we examine its predictions for the dominant modes of a representative KG orbit.
Figure~\ref{fig:top20_comparison} shows a detailed comparison for a sample with parameters $(a, p, e, x_I) = (0.30, 16.26, 0.20, -0.50)$.
For this sample, the model achieves a mode-distribution error of $1.33 \times 10^{-3}$.
The top 20 dominant modes, shown as blue bars, span a full order of magnitude in power.
Our surrogate's predictions (orange bars) closely track the true amplitudes magnitudes across this dynamic range.
For this sample, the Mean Absolute Percentage Error (MAPE) between the predicted and true amplitudes is 3.95\%, demonstrating the model's ability to maintain quantitative accuracy even for sub-dominant modes in a complex spectrum.

In addition to the accuracy of the surrogate model, we also evaluate its computational performance, which is a key factor.
As shown in Table~\ref{tab:inference_benchmark}, on a single NVIDIA GeForce RTX 3090 GPU, the inference time for our trained network is on the order of milliseconds per parameter point. 
The trained model contains about 0.67M FP32 trainable parameters, corresponding to about 2.7 MB of parameter storage. Because the forward graph and output tensor dimensions are fixed, the inference cost is essentially independent of orbital complexity for the present surrogate, and the timing variability is dominated mainly by batch size and normal runtime fluctuations. The runtime memory usage grows by about 9.8 MB per additional sample, consistent with the benchmark values reported in Table~\ref{tab:inference_benchmark}.
The throughput increases significantly with batch size, saturating at approximately 795 samples/sec.
This corresponds to an average inference time of about 1.26 ms per sample at optimal batch sizes.
This millisecond-scale inference time represents a speed-up of several orders of magnitude compared to numerical Teukolsky solvers, which can require hours or even days to compute the amplitude sets for a single parameter point.

To understand the practical implication of this performance for full inspiral waveform generation, we can contextualize it within an adiabatic framework like \ac{FEW}~\cite{Katz:2021yft, Chapman-Bird:2025xtd}.
In such frameworks, a full waveform is constructed by evaluating snapshot amplitudes at a series of sparse points (typically ${\sim 100}$) along the inspiral trajectory.
If we consider that ${\sim 100}$ distinct orbital parameter sets can be processed in a single, parallelized batch, the total wall-time for generating the complete set of amplitude data for one inspiral waveform is expected to be less than one second.

\begin{table}[!htbp]
\small
\caption{Inference performance of the surrogate model on an NVIDIA GeForce RTX 3090 GPU.
The reported throughput and average time per sample are measured with float32 precision.
Peak memory usage scales with the batch size, while throughput saturates at higher batch sizes, achieving a minimum average inference time of approximately 1.26 ms per sample.}
\label{tab:inference_benchmark}
\setlength{\tabcolsep}{4pt} % Adjust column spacing
\begin{ruledtabular}
\begin{tabular}{rD{.}{.}{2}D{.}{.}{3}D{.}{.}{2}}
\multicolumn{1}{c}{Batch} & \multicolumn{1}{c}{Throughput} & \multicolumn{1}{c}{Time/Sample} & \multicolumn{1}{c}{Peak Mem.} \\
\multicolumn{1}{c}{Size} & \multicolumn{1}{c}{(s$^{-1}$)} & \multicolumn{1}{c}{(ms)} & \multicolumn{1}{c}{(MB)} \\
\hline
1 & 112.86 & 8.861 & 21.38 \\
2 & 228.76 & 4.371 & 31.91 \\
4 & 434.58 & 2.301 & 52.56 \\
8 & 522.55 & 1.914 & 92.86 \\
16 & 645.60 & 1.549 & 169.95 \\
32 & 730.16 & 1.370 & 330.60 \\
64 & 780.61 & 1.281 & 647.50 \\
128 & 786.49 & 1.271 & 1277.09 \\
256 & 793.96 & 1.260 & 2546.47 \\
512 & 795.58 & 1.257 & 5070.13 \\
1024 & 792.98 & 1.261 & 10128.78 \\
2048 & 792.24 & 1.262 & 20246.06 \\
\end{tabular}
\end{ruledtabular}
\end{table}

\section{Conclusion} \label{sec:conclusion}
In this work,  we have constructed a convolutional encoder-decoder surrogate that predicts the full Teukolsky amplitude spectrum for generic Kerr EMRIs.
Trained via a curriculum that progresses from Schwarzschild to generic Kerr orbits, the model evaluates in milliseconds with a median mode-distribution error of $\sim 10^{-5}$ for simple orbits and approximately $\sim 10^{-3}$ for generic configurations.

This level of performance should be interpreted in the context of EMRI data analysis requirements.
While high-precision parameter estimation ultimately demands smaller waveform mismatches, a baseline requirement for the amplitude mode-distribution error is on the order of $10^{-2}$ to ensure unbiased signal extraction for a significant fraction of sources~\cite{Isoyama:2021jjd, Chapman-Bird:2025xtd}.
The median error of our model meets this criterion for all orbital types, including the most complex generic configurations.
This performance is achieved despite the challenges posed by the curse of dimensionality and the discontinuous mode coverage in our training data, as detailed in Secs.~\ref{sec:methodology} and \ref{sec:result}.

This proof-of-concept surrogate is not yet ready for integration into a practical adiabatic waveform framework, as its domain of validity is constrained by the PN training data to the weak-field regime ($p \gtrsim 10$) and low eccentricities ($e \le 0.25$).
This restricted parameter space implies not only a smaller domain but also a simpler mode spectrum compared to what is expected from a fully relativistic dataset covering the strong-field, high-eccentricity regime~\cite{Drasco:2005kz, Hughes:2021exa}.

The immediate path forward is therefore to train the model on amplitudes generated by fully relativistic Teukolsky numerical solvers.
This will extend the surrogate's validity into the strong-field and high-eccentricity regimes and across all relevant mode indices.
The current PN-trained model will serve as a strong starting point for this endeavor, with its learned weights providing a robust initialization for transfer learning on sparse and computationally expensive numerical data.

Extending this work to a larger parameter range will present further challenges.
Future improvements in the more difficult sectors will likely require both denser sampling and further methodological refinement. A dedicated scaling analysis is left for future work, especially because the numerical behavior may differ substantially for spherical-basis amplitudes or fully relativistic numerical datasets.
To address this, we plan to employ an iterative, error-driven adaptive sampling scheme to construct the next-generation surrogate.
A preliminary model, trained on a sparse grid, can be used to identify regions of high error, thereby guiding the targeted generation of new training points only where they are most needed.

The surrogate modeling framework developed here is not limited to predicting point-particle Teukolsky amplitudes at infinity; it can also be readily extended to fit the first-order self-force (1SF), thereby enabling the construction of complete adiabatic inspiral waveforms for generic Kerr orbits.
By demonstrating a fast and flexible framework for the EMRI amplitude problem, this work lays a foundation for the fast waveform models required for TianQin and LISA data analysis.

\begin{acknowledgments}

We would like to thank Soichiro Isoyama for helpful discussions and assistance with the PN dataset.
This work is supported by the National Key Research and~Development Program of China (Grant No. 2023YFC2206703, 2021YFC2203002),
and the National Science Foundation of China (Grant No. 12575071).
The main calculations in this paper rely on the \ac{BHPC}.
We also made use of the kerrgeopy package for geodesic calculations.

\end{acknowledgments}

\bibliography{EMRITransfer}
\end{document}